\documentclass[pra,balancelastpage,nofootinbib,twocolumn,fleqn,showkeys]{revtex4}%
\usepackage{amsfonts}
\usepackage{amsmath}
\usepackage{amssymb}
\usepackage{color}
\usepackage{float}
\usepackage{hyperref}
\usepackage{textcomp}
\usepackage{subfigure}
\usepackage[rightcaption]{sidecap}
\usepackage{graphicx}
\begin{document}
\title{Electromagnetically induced transparency and tunable fano resonances in hybrid optomechanics}
\author{Muhammad Javed Akram$^{1}$}
\email{javed\_quaidian@yahoo.com}
\author{Fazal Ghafoor$^{2}$}
\author{Farhan Saif$^{1}$}
\affiliation{$^1$Department of Electronics, Quaid-i-Azam University, 45320 \ Islamabad, Pakistan. \\ $^2$Department of Physics, COMSATS Institute of Information Technology, Islamabad, Pakistan}
\begin{abstract}
We explain the phenomena of electromagnetically induced transparency (EIT) of a weak probe field and tunable Fano resonances in hybrid optomechanics. The system of study consists of a two-level atom coupled to a single-mode field of an optomechanical resonator with a moving mirror. We show that a single EIT window exists in the presence of optomechanical coupling or Jaynes-Cummings coupling, whereas two distinct double EIT windows occur when both the couplings are simultaneously present. Furthermore, based on our analytic and numerical work,
we prove the existence of tunable Fano resonances in the system. The controlling parameters of the system, which switch from a single EIT window to double EIT windows and are needed to tune the Fano resonances, can be realized in present-day laboratory experiments.
\end{abstract}
\maketitle
%________________________________________________________________
\section{Introduction}
Experimental progress in last decade has made cavity optomechanics
a playground for the study of a variety of regimes, from quantum ground-state cooling \cite{1} to strong coupling dynamics \cite{2}. Combining optomechanics with mechanical elements like nano/micro mechanical membranes \cite{3} on one hand, and with atoms \cite{4} and Bose-Einstein condensate (BEC) on the other hand \cite{5}, leads to hybrid systems that serve as workhorses to explore coherent dynamics in microscopic and macroscopic domains \cite{6,7,8}. Electromagnetically induced transparency (EIT) is a direct manifestation of quantum coherence that takes place in a physical system when there are two or more different pathways for a transition \cite{9,10,11}. EIT has not only made valuable contributions in the developments of optical sciences, but has contributed to many advanced applications like slow light \cite{12}, dark-state polariton \cite{13} and induced photon-photon interaction in cold atomic gases \cite{14}. The pump-probe response for the optomechanical systems shares the properties of paradigmatic $\Lambda$-system, and display EIT \cite{15} and double EIT phenomena \cite{16}.

Symmetric Lorentzian and asymmetric Fano line shapes are ubiquitous spectral features, and have appeared as fundamental spectroscopic signatures that quantify the structural and dynamical properties of atoms and molecules \cite{17,18}. Ugo Fano discovered the effect as an interference between resonant and non-resonant processes in photoionization \cite{19}. Since its discovery, the asymmetric Fano resonance has been a characteristic feature of interacting quantum systems, and has been found in various domains, more notably in quantum dots \cite{20}, plasmonic nanoparticles \cite{21}, photonic crystals \cite{22}, and electromagnetic metamaterials \cite{23}. Since EIT results from the interference of different contributions, we can expect Fano-like shapes in EIT under certain conditions \cite{24}. More recently, the Fano resonances have been demonstrated in optomechanics with double-cavity configuration \cite{25}.

Our present work is motivated by the recent work in \cite{7}
and is based on a hybrid quantum system, combining cavity quantum
electrodynamics (QED) \cite{8,26} and optomechanics. In this paper, we report that EIT as well as double EIT windows exist in the hybrid system. In addition, we explain the occurrence of tunable single and double Fano resonances in the system by controlling the cavity parameters. The parametric values of the hybrid system used in our study are realizable in present-day laboratory experiments. Our analytical and numerical results obtained for these parametric values show very good agreement.

The paper is organized as follows: In Sec.~\ref{sec2}, we present the hybrid system based on atom-cavity optomechanics with detailed analytical discussions of the Heisenberg-Langevin equations and standard input-output theory. In Sec.~\ref{sec3}, we demonstrate the phenomenon of electromagnetically induced transparency, where single and double EIT windows have been obtained. In Sec.~\ref{sec4}, we explain the necessary conditions that lead to single and double Fano resonances in the hybrid system. Finally in Sec.~\ref{sec5}, we summarize the results.
%%%%%%%%%%%%%%%%%%%%%%%%%%%%%%%%%%%%%%%%%%%%%%%%%%%%%%%%%%%%%%%%%%%%%%%%%%%%%%
\section{The Model System}\label{sec2}
\begin{figure}[ht]
\includegraphics[scale=0.6]{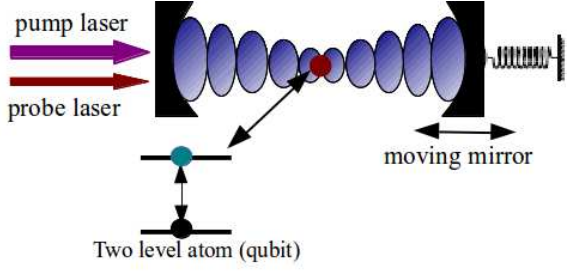}
\caption{Schematic representation of our hybrid system: A strong driving field of frequency, $\omega_l$, and a weak probe field of frequency, $\omega_p$, are injected in the cavity. The cavity field is coupled to a moving mirror and a two-level atom, forming a hybrid optomechanical set-up.} \label{fig1}
\end{figure}

We consider a hybrid cavity QED-optomechanical system where a single cavity field mode is coupled both to a mechanical resonator and to a two-level atom, as shown in figure \ref{fig1}. Thus, the system has the usual nonlinear coupling of optomechanics and  Jaynes-Cummings (JC) coupling of cavity QED architectures \cite{8}.
We assume that the cavity is simultaneously driven by a strong pump-field of frequency $\omega_l$ and a weak probe-field of frequency $\omega_p$ respectively. Hence, the total Hamiltonian of the system can be written as:
\begin{equation}
H_T= H_{m}+H_{c}+H_{a}+H_{int}+H_{lp},
%H_{mech}&=&\dfrac{p^2}{2m} + \dfrac{1}{2} m \omega_m^2 q^2 \\
%H_{opt}&=&  \hbar \omega_c c^\dagger c \\
%H_{int}&=& -\hbar g_{cm} c^{\dagger}c q + \hbar g_{ac}(c^\dagger\sigma_+ + c %\sigma_-), 
\end{equation}
where,
\begin{subequations}
\begin{align}
H_{m}&=\dfrac{p^2}{2m} + \dfrac{1}{2} m \omega_m^2 q^2 \label{Hm},\\
H_{c}&= \hbar \omega_c c^\dagger c \label{Hc}, \\
H_{a}&=\hbar \dfrac{\omega_a}{2}\sigma_z \label{Ha},\\
H_{int}&= - \hbar g_{mc}c^{\dagger}c q + \hbar g_{ac}(c^\dagger\sigma_+ + c \sigma_-)\label{Hint}, \\
H_{lp}&=i\hbar(\Omega_l e^{-i\omega_lt}c^\dagger - \Omega_l^* e^{i\omega_lt}c) \nonumber \\ &+ i\hbar(\varepsilon_p e^{-i\omega_pt}c^\dagger - \varepsilon^*_p e^{i\omega_pt}c).\label{l-p}
\end{align}
\end{subequations}
Here, $H_m$ gives the free Hamiltonian of the moving mirror. The parameters $q$ and $p$, represent the position and momentum operators of the mechanical resonator with the vibration frequency $\omega_m$ and the mass $m$. Furthermore, $H_c$ describes the Hamiltonian of the cavity field with the creation (annihilation) operator $c^\dagger$ ($c$) of the single mode of cavity field; $H_a$ accounts for Hamiltonian of the two-level atomic system, or qubit, with transition frequency $\omega_a$. $H_{int}$ describes the interaction Hamiltonian, where first term shows the interaction between the cavity and the oscillating mirror with optomechanical coupling $g_{mc}$, and the second term denotes the atom-field (Jaynes-Cummings) coupling $g_{ac}$. The parameters $\sigma_+$ and $\sigma_-$ are the raising and lowering operators of the two level atom with transition frequency $\omega_a$ and are expressed by $\sigma_\pm=\frac{\sigma_x \pm i\sigma_y}{2}$. Here $\sigma_{x,y,z}$ are the Pauli spin operators for the two-level atom. Finally, the last part $H_{lp}$ corresponds to the classical fields, pump and probe lasers, with frequencies $\omega_l$ and $\omega_p$ respectively. Moreover, $\Omega_l$ and $\varepsilon_p$ are related to the laser power $P$ by $|\Omega_l|=\sqrt{2\kappa P_{l}/ \hbar \omega_l}$ and $\varepsilon_p=\sqrt{2\kappa P_{p}/ \hbar \omega_p}$ respectively, with $\vert\Omega_l\vert >> \vert\varepsilon_p\vert$.
In the rotating reference frame at the frequency $\omega_l$ of the strong driving field, the combined Hamiltonian of the system can be written as:
\begin{eqnarray}
H_T &=& \hbar \Delta_c c^{\dagger}c + \dfrac{p^2}{2m} + \dfrac{1}{2} m \omega_m^2 q^2 + \hbar \dfrac{\Delta_a}{2}\sigma_z - \hbar g_{mc}c^{\dagger}cq  \nonumber \\ &+&
\hbar g_{ac}(c^\dagger\sigma_+ +c\sigma_-) + i\hbar(\Omega_l c^\dagger -\Omega_l^* c) \nonumber \\ &+&
i\hbar(\varepsilon_p e^{-i\Delta t}c^\dagger - \varepsilon_p^* e^{i\Delta t}c),\label{Ham2}
\end{eqnarray}
where, $\Delta_c=\omega_c-\omega_l$, $\Delta_a=\omega_a-\omega_l$ and $\Delta=\omega_p-\omega_l$ are the respective detunings of the cavity field, two-level atom and the probe field with respect to driving field frequency $\omega_l$.

For a detailed analysis of the system, we include photon losses in the cavity and the Brownian noise acting on the mirror as well as the decays associated with the atom. Based on the Hamiltonian in Eq.~(\ref{Ham2}), the quantum Langevin equations (QLE) can be written as
\begin{subequations}
\begin{align}
\dot{q}&=\dfrac{p}{m}, \\
\dot{p}&=-m\omega_m^2 q - \gamma_m p + g_{mc}c^\dagger c + \xi(t), \\
\dot{c}&=-(\kappa+i\Delta_c)c+i g_{mc} c q-ig_{ac}\sigma_- +\Omega_l \\ &+\varepsilon_p e^{-i\Delta t} + \sqrt{2\kappa}c_{in}(t), \nonumber\\
\dot{\sigma}_-&=-(\gamma_a+i\Delta_a)\sigma_- + ig_{ac} c \sigma_z+ \sqrt{2\gamma_a}a_{in}(t).
\end{align}
\end{subequations}
Here, we have introduced the input vacuum noises associated with the cavity field $c_{in}(t)$ and the atomic system $a_{in}(t)$ respectively, having zero mean values. Moreover, $\kappa$, $\gamma_a$ and  $\gamma_m$ denote the radiative decays associated with the cavity, the atom,  and the mechanical mode respectively. The input vacuum noises affecting cavity field and the two-level atom, obey the non-vanishing commutation relations \cite{27}, that is, $\langle c_{in}(t) c_{in}^\dagger (t') \rangle = \delta(t-t')$ and  $\langle a_{in}(t) a_{in}^\dagger (t') \rangle = \delta(t-t')$. The Hermitian Brownian noise operator (thermal Langevin force) $\xi(t)$ with zero mean value $\langle \xi(t) \rangle =0$ has correlation function \cite{28}:
\begin{equation}
\langle \xi(t) \xi^\dagger (t') \rangle = \int \omega e^{-i\omega (t-t')}N(\omega)d\omega,
\end{equation}
where, $N(\omega)=\dfrac{\gamma_m}{2\pi\omega_m} \left[ 1+coth \right( \dfrac{\hbar \omega}{2k_B T} \left) \right] $. Here, $k_B$ is the Boltzmann constant and $T$ is the temperature of the mechanical oscillator reservoir.

We obtain steady-state solutions of the system operators and study the output spectrum by using the mean field approximation (factorization assumption) \cite{15}, viz, $\langle qc\rangle =\langle q\rangle \langle c\rangle $. We write the time evolutions of the expectation values of the system operators as
\begin{subequations}\label{meq}
\begin{align}
\dfrac{d\langle p \rangle}{dt}&=-m\omega_m^2 \langle q \rangle - \gamma_m \langle p \rangle + g_{mc}\langle c^\dagger \rangle \langle c \rangle, \\
\dfrac{d\langle q \rangle}{dt}&=\dfrac{\langle p \rangle}{m}, \\
\dfrac{d\langle c \rangle}{dt}&=-(\kappa+i\Delta_c) \langle c  \rangle +i g_{mc}  \langle c \rangle \langle q  \rangle -ig_{ac} \langle \sigma_- \rangle \nonumber\\ &+\Omega_l + \varepsilon_p e^{-i\Delta t}, \\
\dfrac{d\langle \sigma_- \rangle}{dt}&=-(\gamma_a+i\Delta_a) \langle \sigma_- \rangle  + ig_{ac}  \langle c \rangle  \langle \sigma_z  \rangle.
\end{align}
\end{subequations}
In obtaining the expectation values of the operators in the above set of equations, we drop the Hermitian Brownian noise and input vacuum  noise terms which are averaged to zero. Moreover, the above nonlinear equations cannot be solved exactly, because the steady-state response contains an infinite number of components of different frequencies of the coupled system. In order to obtain the steady-state solutions which are exact for the strong driving $\Omega_l$ and to the lowest order in the weak probe $\varepsilon_p$, we make the following ansatz \cite{29}:
\begin{subequations}
\begin{align}
\langle c \rangle = c_s + c_- e^{-i\Delta t} + c_+ e^{i\Delta t}, \label{cmean} \\
\langle q \rangle=q_s + q_- e^{-i\Delta t} + q_+ e^{i\Delta t}, \\
\langle \sigma_- \rangle=a_s + a_- e^{-i\Delta t} + a_+ e^{i\Delta t}, 
\end{align}
\end{subequations}
where, $c_s$, $q_s$ and $a_s$ are the steady-state solutions when $\varepsilon_p=0$. Moreover, $c_\pm$ are much smaller than $c_s$ and are of the same order as $\varepsilon_p$. Similarly for $\langle q \rangle$ and $\langle \sigma_- \rangle$. By using the above ansatz, we obtain the following steady-state solutions
\begin{equation}\label{cs}
c_s=\dfrac{\Omega_l}{\kappa + i\tilde{\Delta} - \frac{g_{ac}^2 \langle \sigma_z \rangle_{ss}}{(\gamma_a+i\Delta_a)}},
\end{equation}
\begin{equation}\label{cm}
c_-=\dfrac{\varepsilon_p (A-B)}{B B' + (A-C)(A'+C)- (AB' +  A' B) + 2iC\tilde{\Delta}},
\end{equation}
where,
\begin{subequations}\label{B}
\begin{align}
A &=\kappa-i\Delta_c-i\tilde{\Delta} + \dfrac{i g_{mc}^2}{m\hbar(\omega_m^2-i\gamma_m \Delta-\Delta^2)}\vert c_s \vert^2, \\
B &=\dfrac{g_{ac}^2\langle \sigma_z \rangle_{ss}}{\gamma_a-i\Delta_a-i\Delta}, \label{10b} \\
C &=\dfrac{i g_{mc}^2}{m\hbar(\omega_m^2-i\gamma_m \Delta-\Delta^2)}\vert c_s \vert^2, \\
\tilde{\Delta} &=\Delta_c-\dfrac{g_{mc}^2}{m\hbar \omega_m^2}\vert c_s \vert^2.
\end{align}
\end{subequations}
Here, $A'=(A(-\Delta))^*$ and $B'=(B(-\Delta))^*$. The expressions (\ref{cs}) and (\ref{cm}) assist us to study the output field at the probe frequency. Moreover, here we do not mention the expression for $c_+$ as this describes the four-wave mixing for the driving field and the weak probe field (which shall be discussed separately). Since, we are interested in the mean response of the system to the probe field, while there is a large detuning between the cavity field and the atomic system. Without loss of generality,
hereafter we assume that the atom stays in its excited state for a long time, which implies that its steady-state value is set as $\langle \sigma_z \rangle_{ss}=1$ \cite{30}. Furthermore, we assume that the system operates in the sideband resolved regime i.e. $\omega_m \gg \kappa$ and $\Delta \sim \omega_m$ \cite{15}.

The response of the system to all frequencies can be detected by the output field, which can be obtained via standard input-output theory \cite{31},
\begin{equation}
\langle c_{out}(t) \rangle  + \dfrac{\Omega_l}{\sqrt{2\kappa}} + \dfrac{\varepsilon_p}{\sqrt{2\kappa}} e^{-i\Delta t}=\sqrt{2\kappa}\langle c \rangle. \label{in-out}
\end{equation}
We can express the mean value of the output field from Eqs.~ (\ref{cmean}) and (\ref{in-out}) as
%\begin{widetext}
\begin{eqnarray}
\sqrt{2\kappa}\langle c_{out}(t) \rangle&=&(2\kappa c_s - \Omega_l) + (2\kappa \dfrac{c_-}{\varepsilon_p} - 1)\varepsilon_p e^{-i\Delta t} \nonumber \\
&+& 2\kappa (\dfrac{c_+}{\varepsilon_p^*})\varepsilon_p^* e^{i\Delta t}. \label{out}
\end{eqnarray}
%\end{widetext}
Note that, the second term on the right side in the above expression corresponds to the response of the whole system to the weak probe field at frequency $\omega_p$ via the detuning $\Delta=\omega_p - \omega_l$. Hence, the real and imaginary parts of the amplitude of this term accounts for the absorption
and dispersion of the whole system to the probe field, or equivalently the properties of the medium inside the cavity. We write the amplitude of the rescaled output field corresponding to the weak probe field as
\begin{equation}
E_{out} = \dfrac{2}{\varepsilon_p} \kappa c_-. \label{eout}
\end{equation}
The corresponding real and imaginary parts of the output probe field are, $Re[E_{out}]=\frac{\kappa(c_- + c_-^*)}{\varepsilon_p}$, and $Im[E_{out}]=\frac{\kappa(c_- - c_-^*)}{i\varepsilon_p}$, which correspond to absorption and dispersion respectively. These two quadratures of the output field can be measured by homodyne detections \cite{31}.
%****************************************************************************
\section{From Single to double EIT window}\label{sec3}
\begin{figure}[t]
\includegraphics[width=0.45\textwidth]{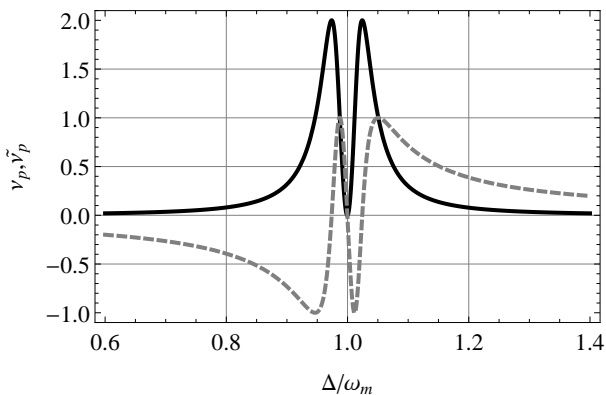}
\caption{Absorption, $\nu_p$,  and dispersion, $\tilde{\nu}_p$, profiles are shown versus the normalized probe field detuning. In the absence of atom-field coupling, i.e. $g_{ac}=0$, we find single EIT window for non-zero optomechanical coupling. Experimental parameters are set as: $\Omega_l/2\pi=10$ MHz, $g_{mc}/2\pi=2$ MHz, $\omega_m /2\pi=100$ MHz, $\Delta_c/2\pi=100$ MHz, $\kappa/2\pi=4$ MHz and $\gamma_m/2\pi=1$ KHz.} \label{EIT1}
\end{figure}

In this section, we expalin the EIT and double EIT phenomena which emerges due to the interaction between the cavity field and the mechanical mode and due to the additional atom-field coupling in the hybrid system. We write the output field at the probe frequency as
\begin{equation}
E_{out}=\nu_p + i \tilde{\nu_p}. \label{Eout}
\end{equation}
Here, the real and imaginary parts, $\nu_p$ and $\tilde{\nu_p}$ respectively, account for the inphase and out of phase quadratures of the output field at probe frequency and correspond to the absorption and dispersion as defined above. In order to demonstrate EIT in the system, we consider the parameters from the experiments reported in \cite{32,33}. The optomechanical coupling $g_{mc}/2\pi=4$ MHz, $g_{ac}/2\pi=10$ MHz, $\Omega_l/2\pi=20$ MHz, $\omega_m /2\pi=100$ MHz, $\Delta_c/2\pi=100$ MHz, $\Delta_a/2\pi=30$ MHz, $\gamma_a/2\pi=0.01$ MHz, $\kappa/2\pi=4$ MHz and $\gamma_m/2\pi=147$ Hz.
In Fig.~\ref{EIT1}, we show the phase quadratures $\nu_p$ (black-solid line) and $\tilde{\nu}_p$ (gray-dashed line) as a function of the normalized detuning $\Delta/ \omega_m$. In the presence of optomechanical coupling ($g_{mc}\neq0$) and absence of atom-field coupling ($g_{ac}=0$), we observe the generic EIT window \cite{15}. In addition to the coupling parameter $g_{mc}$, the EIT window in Fig.~\ref{EIT1}, can be modulated effectively by the power of the pump laser, or equivalently by $\Omega_l$ \cite{34,10}. On increasing $\Omega_l$ or $g_{mc}$, the window expands or vice versa. In what follows, in the limit of small $\Omega_l$ the $c_s$ vanishes, and the two peaks in Fig.~\ref{EIT1} converge to a single peak, making a standard Lorentzian absorption peak with a full width at $\Delta=\omega_m$ \cite{10,16}. However, in the presence of strong pump laser, the single Lorentzian peak splits into two peaks, and the two peaks are further and further apart if we continuously increase $\Omega_l$ (or $g_{mc}$).

\begin{figure}[t]
\includegraphics[width=0.45\textwidth]{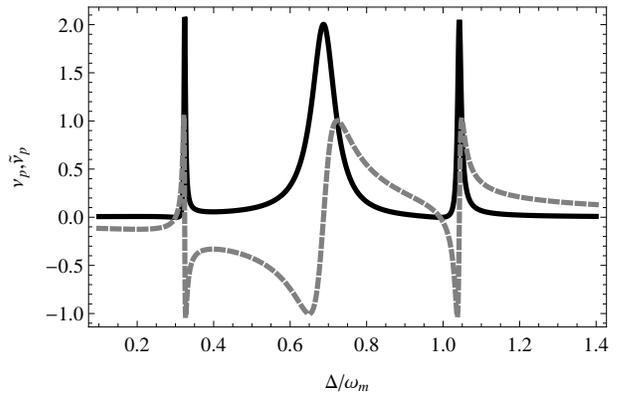}
\caption{Absorption and dispersion profiles are shown versus the normalized probe detuning. Double-EIT features are observed in the presence of both coupling parameters. Here, $g_{ac}/2\pi=10$ MHz, $g_{mc}/2\pi=2$ MHz and $\Omega_l/2\pi=20$ MHz. Parameters for atomic system are: $\Delta_a/2\pi=40$ MHz, $\gamma_a/2\pi=0.01$ MHz. Rest of the parameters are the same as in Fig.~\ref{EIT1}.} \label{EIT2}
\end{figure}

On the other hand, as we switch on the atom-field coupling $g_{ac}$ along with $g_{mc}$, we observe that the resonant character of the weak probe field changes from single EIT window to double EIT window; as shown in Fig.~\ref{EIT2}.
The splitting of the single EIT window into double EIT window as a function of the Jaynes-Cummings coupling $g_{ac}$ is a consequence of the formation of the dressed state mediated by the single-photon state and the two-level atom \cite{8}. Interestingly, double EIT window can be transformed to single EIT window if optomechanical coupling is gradually switched off. Hence, we can control and tune the EIT window by controlling the Jaynes-Cummings coupling $g_{ac}$, as presented in Fig.~\ref{gfano}. 

Splitting of single EIT window to two distinct EIT windows can be understood analytically as we examine the structure of the output field. In the case of single EIT window, when $g_{ac}=0$, Eq.~(\ref{eout}) leads to the output field $E_{out}$, 
\begin{equation}
E_{out}=\dfrac{2\kappa}{(\kappa-i(\Delta-\omega_m)+\frac{\omega_m(\Delta_c-\tilde{\Delta})/2}{\gamma_m/2-i(\Delta-\omega_m)}}. \label{eout1}
\end{equation}
as obtained in Ref.~\cite{15}. The denominator of Eq.~(\ref{eout}) has two roots which are,
\begin{equation}
x_{1,2}=-\dfrac{i(2\kappa+\gamma_m)}{4} \pm \dfrac{1}{2}\sqrt{-(\kappa-\gamma_m/2)^2 -2\omega_m(\tilde{\Delta}-\Delta_c)}
\end{equation}
Here, $x=\Delta-\omega_m$ accounts for the detuning from the line center. These two roots provide the location of the two maxima for a single EIT window as shown in Fig.~\ref{EIT1}. In the presence of the atom-field coupling $g_{ac}$ in the system, the structure of the output field changes that results in the splitting of EIT from single to double. In this case, in the denominator of Eq.~(\ref{eout}), we find additional terms with $\kappa$ as a consequence of non-zero $g_{ac}$.
Keeping in view the structure of Eqs.~(8-10), we note that $c_s$ and thereafter $c_-$ are modified. Hence, for the case of $\Delta_a=\omega_m=\Delta-x$, the denominator of the output field becomes cubic. 
Unlike in the case of single cavity \cite{15}, there exists three roots in the present case. The third root occurs due to the presence of atom in the hybrid system. Therefore, we found three peaks in the absorption profile, as in Fig.~\ref{EIT2}. Here each of the three roots depicts a maximum in double EIT window. Thus, analytically the structure of the output field confirms the splitting of single EIT window into two distinct double EIT windows. The two windows occur for large enough values of $g_{ac}$. In addition, the atom contributes in the destructive interference, since $\omega_a \neq \omega_l$. Hence, the interference between four parties, namely; atom, moving mirror, pump and probe fields respectively, corresponding to $\omega_a$, $\omega_m$, $\omega_l$ and $\omega_p$, causes the splitting of single EIT window, and emergence of two distinct EIT windows. Therefore, dispersion in the output field changes from normal to anomalous. It is also worth noticing that analogous to the multi-level atomic systems, single-ended optomechanical systems share the properties of three-level atomic system, which leads to the occurrence of single EIT window \cite{15}. The occurrence of double EIT window requires an additional interfering pathway, and the hybrid optomechanical systems spectacularly meet this requirement, as they can share the properties of four (or more) level atomic systems \cite{36}.
%__________________________________________________________________________
\section{Fano Resonances in the Hybrid System}\label{sec4}
Ugo Fano \cite{19} in his quantum mechanical study of the autoionizing states of atoms, discovered that the quantum interference of different transition amplitudes leads to absorption profile which exhibits minima or zero. Thus, the asymmetric Fano resonance has been a characteristic feature of interacting quantum systems, and the shape of this resonance is distinctively different in contrast of conventional symmetric resonance curves like a Lorentzian resonance or the symmetry in EIT windows \cite{17,25}. At grass-roots level, the origin of the Fano resonance is in the constructive and destructive interference of a narrow discrete resonance with a broad spectral line or a continuum. Since EIT results from the interference of different contributions, we can observe Fano line shapes in EIT by suitably tuning the system parameters \cite{24,25}. Following, we obtain the Fano profiles for each of the coupling parameters. Simultaneous presence of both coupling parameters causes the splitting of single Fano resonances into doubly Fano resonances in the hybrid system.
%__________________________________________________________________________
\subsection{Fano Resonance in the output field}
\begin{figure}[t]
\includegraphics[width=0.45\textwidth]{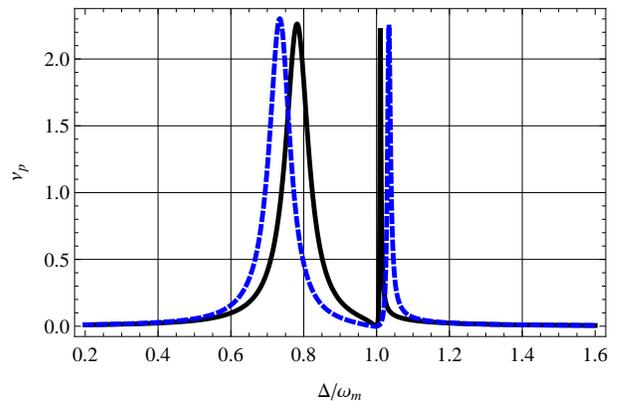} 
\caption{Fano line shapes in the absorption profile are shown for $g_{mc}/2\pi=2$ MHz (blue dashed-curve) and $g_{mc}/2\pi=4$ MHz (black solid-curve). Here, $\Omega_l/2\pi=20$ MHz and $\Delta_c/2 \pi=80$ MHz. All other parameters are the same as in Fig.~(\ref{EIT1}). }\label{fano1}
\end{figure}

In order to demonstrate the emergence of Fano resonances in the hybrid system, we show the absorption profiles of the output field at probe frequency with certain range of controlling parameters. We first consider the case of single EIT window, i.e. when atom-field coupling is off and only optomechanical coupling is present. From Eq.~(\ref{eout1}), we obtain the following Fano profile \cite{25},
\begin{equation}\label{fanno1}
\nu_p=Re[E_{out}]=\dfrac{2}{1+q_1^2}\dfrac{(x_1+q_1)^2}{1+x_1^2},
\end{equation}
where $x_1 = \frac{\Delta-\omega_m}{\Gamma_1} -q_1$, $\Gamma_1=\frac{\kappa\beta}{\kappa^2+\Omega^2}$, $q_1 =-\Omega/\kappa$, $\beta=\omega_m(\Delta_c-\tilde{\Delta})/2$ and $\Omega=\Delta_c-\omega_m$. Note that, the absorption profile in Eq.~(\ref{fanno1}) has exactly the same form as the classic profile of Fano resonance with maximum at $x_1 = 1/q_1$ and zero at $x_1 = -q_1$. The asymmetry parameter $q_1$ is related to the frequency offset $\Omega$. Here, the frequency offset factor $\Omega$ plays an important role in the emergence of the asymmetric Fano shapes. Physically it means that the anti-Stokes process is not resonant with the cavity frequency. In Fig.~\ref{fano1}, we show that when detuning of the cavity field $\Delta_c$ is out of resonance with mechanical frequency, for instance $\Delta_c < \omega_m$, consequently Fano shapes appear in EIT spectra. In Fig.~\ref{fano1}, we show a typical behavior of absorption for $\omega_m=100$ MHz and $\Delta_c=80$ MHz for different values of the mirror-field coupling $g_{mc}$. Fig.~\ref{fano1} indicates that both absorption profiles have common minima or zero, which confirms that these profiles are Fano resonances \cite{24}.

\begin{figure}[t]
\includegraphics[width=0.44\textwidth]{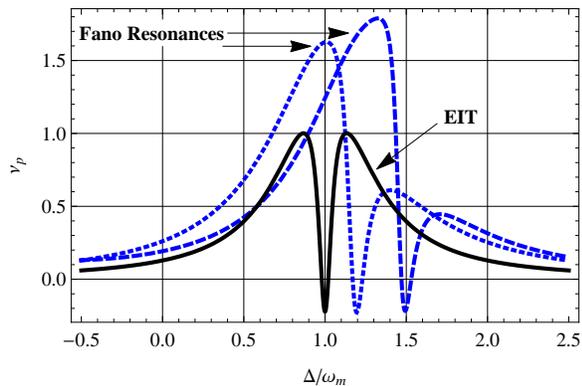}
\caption{Fano resonances in the absorption profile are shown in the absence of coupling, $g_{mc}$. The parameters are: $g_{ac}/2\pi=2$ MHz, $\omega_m/2\pi=10$ MHz and $\Omega_l=2$ MHz. The black solid line at $\Delta_a/2\pi=\Delta_c/2\pi=15$ MHz, displays the EIT curve. Fano resonances emerge when $\Delta_a\neq \Delta_c$, as shown for $\Delta_a/2\pi=10$ MHz and $\Delta_c/2\pi=12$ MHz (Blue-dotted curve) and for  $\Delta_a/2\pi=12$ MHz and $\Delta_c/2\pi=15$ MHz (Blue-dashed curve). All other parameters are the same as in Fig.~\ref{EIT1}.}\label{gfano}
\end{figure}

In the next case, we discuss another possibility of Fano resonance, when optomechanical coupling is absent and one can tune the Fano resonances with atom-field coupling. In Fig.~\ref{gfano}, it is shown that the Fano profiles with relatively broad resonance, can be observed near $\Delta_a \sim \omega_m$. For detuning $\Delta_a=\omega_m$ (black curve in Fig.~\ref{gfano}), we obtain the standard EIT profile. Analytically, for the case of $g_{mc}=0$ and $g_{ac}\neq0$, we obtain the following Fano profile (similar to the Fano profile as in Eq.~(\ref{fanno1})), 
\begin{equation}\label{fanno2}
\nu_p=Re[E_{out}]=\dfrac{2}{1+q_2^2}\dfrac{(x_2+q_2)^2}{1+x_2^2}.
\end{equation}
Here, $x_2 = \frac{\Delta-\omega_m}{\Gamma_2} -q_2$, $\Gamma_2=\frac{\gamma_a g_{ac}^2}{\gamma^2+\Delta_a^2}$, and $q_2 =-\Delta_a/\gamma_a$. Hence, in the atom-cavity system, Fano resonances can be tuned with either of the coupling parameters $g_{mc}$ or $g_{ac}$. Each of these Fano line shapes, in Fig.~\ref{fano1} and \ref{gfano}, have a zero point exactly at $x_i = -q_i$. Moreover, our approximation formulae (\ref{fanno1}) and (\ref{fanno2}) and the numerical curves obtained directly from (\ref{eout}) agree. The physical origin of these resonances is related to the constructive and destructive interferences between the weak probe laser in the cavity which can interfere with anti-Stokes sidebands of the stronger pump field induced by the mechanical oscillations or the atomic system. 

Note that, in general, in optomechanical systems we have two coherent processes leading to the building up of the cavity field: (i) the direct building up due to the application of strong pump and weak probe field, and (ii) the building up due to the two successive nonlinear frequency conversion processes between two optical modes and a mechanical mode \cite{25}. These two paths contribute in the interference which leads to the emergence of Fano profiles in the hybrid system. Moreover, EIT occurs when system is in complete resonance, that is, $\Delta=\omega_m$. However due to the present non-resonant interactions, i.e. $\Delta_c\neq \omega_m$ and $\Delta_a\neq \omega_m$, asymmetric Fano profiles appear.
%_________________________________________________________________________
\subsection{Double Fano Resonances}
In this section, we discuss the possibility of double Fano resonances under the combined effect of the coupling parameters $g_{mc}$ and $g_{ac}$, making use of hybrid system to tune the Fano resonances. As we explained in the previous section, due to the existence of the atomic coupling in the system, the structure of the output field changes. When both coupling parameters contribute in the hybrid system, the denominator of Eq.~(\ref{eout}) becomes cubic and hence the number of roots increases from two to three. This is because we have three coupled systems: the cavity mode, the mechanical mode and the two level atom. At the same time, the numerator in (\ref{eout}) becomes a quadratic function of $x$ suggesting the possibility of two different minima in the output field. Therefore, a single Fano resonance splits into a double Fano resonance. 

\begin{figure}[h]
\includegraphics[width=0.44\textwidth]{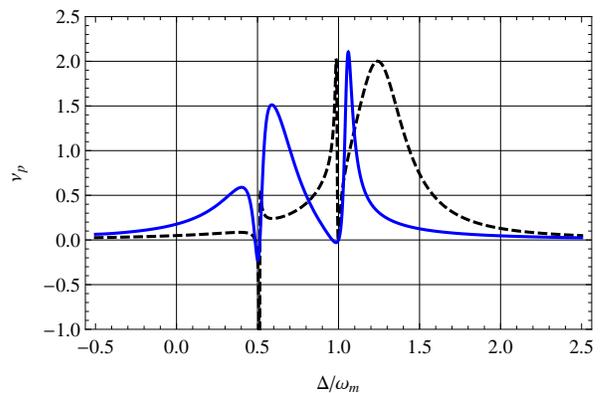}
\caption{The absorption profiles versus the normalized detuning $\Delta/\omega_m$ show double Fano resonances in the output field. Where, Black (dashed) curve corresponds to $g_{mc}/2\pi=0.5$ MHz, $\Delta_c/2\pi=25$ MHz and Blue (solid) curve for $g_{mc}/2\pi=1$ MHz, $\Delta_c/2\pi=15$. Other parameters are set as \cite{33,35}: $g_{ac}/2\pi=2$ MHz, $\omega_m=20$ MHz, $\Delta_a/2\pi=10$ MHz and $\Omega_l/2\pi=2$ MHz. Rest of the parameters are the same as in Fig.~\ref{EIT1}.}\label{dfano2}
\end{figure}

In order to examine the quantitative features of the double Fano resonances, we fix some parameters and tune the others as the parameter space is large enough. In the first case, we fix the atom-field coupling $g_{ac}/2\pi=2$ MHz and the corresponding detuning $\Delta_a/2\pi=10$ MHz, whereas the mirror frequency is set to be $\omega_m/2\pi=20$ MHz. In Fig.~\ref{dfano2}, we show that the prominent feature of double Fano resonances can be observed in the hybrid system by suitably tuning the system parameters.
The two Fano asymmetry parameters, $q_1$ and $q_2$, contribute together in the emergence of double minima in the Fano profiles obtained in Fig~\ref{dfano2}, when both coupling parameters are non-zero. The three coupled systems (the cavity mode, the mechanical mode and the two level atom) provide the three coherent routes for building up the cavity field, which can interfere with each other. As a consequence, a single Fano resonance goes over to a double Fano resonance.

\begin{figure}[!t]
\includegraphics[width=0.45\textwidth]{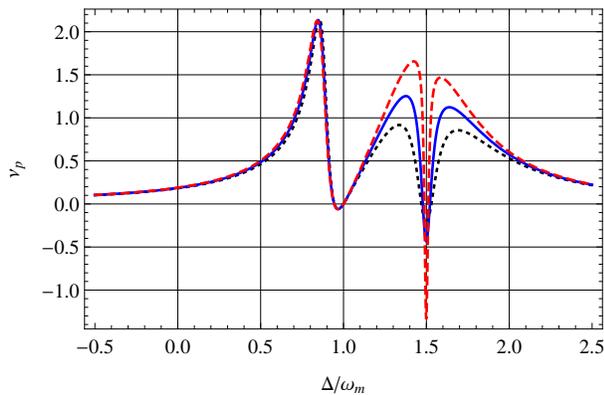}
\caption{Double Fano resonances in the absorption profile are shown. Here, Red (dashed) curve corresponds to $g_{ac}/2\pi=1$ MHz, Blue (solid) for $g_{ac}/2\pi=1.5$ MHz and Black (dotted) for $g_{ac}/2\pi=2$ MHz. Other parameters are set as \cite{33,35}: $g_{mc}/2\pi=0.5$ MHz, $\omega_m=10$ MHz, $\Delta_c=\Delta_a=15$ MHz. Rest of the parameters are the same as in Fig.~\ref{EIT1}.}\label{fano2}
%\end{SCfigure*}
\end{figure}

In Fig.~\ref{fano2}, we fix the optomechanical coupling $g_{mc}$ and tune the Fano resonances with atom-field coupling $g_{ac}$. Combined effect of optomechanical coupling and atom-field coupling produces double Fano minima due to interfering pathways via three subsystems mentioned above. Note that, by increasing the atom-field coupling $g_{ac}$, dip of the lower minima (at right side) increases linearly with increase in $g_{ac}$. In the presence of both coupling, we find the two roots of the nominator of the output field: $\overline{x}_1=-i\frac{\gamma_m}{2}$ and $\overline{x}_2=\frac{1}{2}(-i\gamma_a + \Delta_a - 2 \omega_m)$, which are independent. As we increase any of the coupling parameters, the Fano resonance splits and the splitting increases linearly as we increase the coupling strength. Apart from the splitting, their resonance frequency center is shifted by an amount $\Delta_a/2 -\omega_m$. The frequency splitting is independent of the cavity detuning $\Delta_c$, as long as it is close to $\omega_m$. Therefore, we can always obtain the coupling strength, as well as the coupling power, by measuring the double Fano resonances. If anyone of the coupling is absent, the double Fano resonances goes over to single Fano resonance. Moreover, by using the full features of the hybrid system, we should be able to obtain a lower minimum and a higher maximum in the double Fano resonances as shown in Fig.~\ref{dfano2} and \ref{fano2}. This is analogous to the result in the context of photoionization in which the value of the minimum depends on the radiative effects \cite{25}.

Thus, the results in Fig.~\ref{dfano2} and \ref{fano2} reveal that when the two coupling parameters contribute together, one can observe the double Fano resonances in the hybrid optomechanical system.
Moreover, the presence of a sharp resonance in these absorption profiles, whose asymmetric shapes are consistent with Fano resonances, is remarkably different from the more common Lorentzian resonance \cite{19,18} and symmetric EIT profiles. Fano resonances are able to confine light more efficiently and are characterized by a steeper dispersion than conventional Lorentzian resonances \cite{21,18}, which make them promising for local refractive index sensing applications \cite{23} or surface enhanced Raman scattering \cite{37}.
%%%%%%%%%%%%%%%%%%%%%%%%%%%%%%%%%%%%%%%%%%%%%%%%%%%%%%%%%%%%%%%%%%%%%%%%%
\section{Conclusion}\label{sec5}
In conclusion, we have studied electromagnetically induced transparency and Fano resonances with hybrid atom-cavity optomechanics. In the present contribution, we provide a full analytical model to study the absorption and dispersion profiles of the output field at the probe frequency. It is shown that: (i) in the absence of the coupling between a two-level atom and the cavity field, the EIT window is obtained as noted earlier \cite{15}. (ii) The simultaneous presence of both coupling parameters leads to the double EIT phenomenon in the hybrid system. (iii) We report the prominent signatures of Fano resonances under a wide range of parameters. The Fano resonances are tunable with optomechanical coupling or atom-field coupling. (iv) The combined effects of the optomechanical
and atom-field coupling parameters give rise to the tunable double Fano resonances in the hybrid system. EIT has potential applications in optical switching and slow and fast light effects. However, Fano resonances are able to
confine light more efficiently and are characterized by a steeper dispersion than conventional Lorentzian resonances, making them promising for local refractive index sensing applications or surface-enhanced Raman scattering. Moreover, the set of parameters used in our numerics corresponds to present-day laboratory experiments.
%_____________________________________________________________________
%*******************************************************************************************************


\begin{thebibliography}{99}
\bibitem{30} The value of $\langle \sigma_z \rangle_{ss}$ ranges from $0$ to $1$, where $\langle \sigma_z \rangle_{ss}=$ 1 (0) means the atom in its excited (ground) state. Thus, $\langle \sigma_z \rangle_{ss}$ is a number which can rescale the value of $g_{ac}$, as seen in equations (\ref{cs}) and (\ref{10b}), and it does not affect the occurrence of double EIT window until it is non-zero.
\end{thebibliography}
\end{document}